\documentclass{article}

\usepackage{arxiv}
\usepackage{amsmath}
\usepackage{hyperref}       
\usepackage{url}            
\usepackage{booktabs}       
\usepackage{amsfonts}       
\usepackage{nicefrac}       
\usepackage{microtype}      
\usepackage{graphicx}
\graphicspath{{./images/}}

\usepackage[utf8]{inputenc} 
\usepackage[T1]{fontenc}    
\usepackage[english]{babel} 

\usepackage{xurl}
\usepackage{svg}
\usepackage{xcolor}
\usepackage{longtable}
\usepackage{parskip}
\title{Bluesky: Network Topology, Polarization, and Algorithmic Curation}

\author{%
  Dorian Quelle\\
  Department of Mathematical Modeling and Machine Learning, University of Zurich, Switzerland\\
  Digital Society Initiative, University of Zurich, Switzerland\\
  \texttt{dorian.quelle@uzh.ch}
  \and
  Alexandre Bovet\\
  Department of Mathematical Modeling and Machine Learning, University of Zurich, Switzerland\\
  Digital Society Initiative, University of Zurich, Switzerland\\
}

\begin{document}
\maketitle
\begin{abstract}
Bluesky is a nascent ``Twitter-like'' and decentralized social media network with novel features and unprecedented data access. This paper provides a characterization of its interaction network, studying the political leaning, polarization, network structure, and algorithmic curation mechanisms of five million users. The dataset spans from the website's first release in February of 2023 to May of 2024. 
We investigate the replies, likes, reposts, and follows layers of the Bluesky network. We find that all networks are characterized by heavy-tailed distributions, high clustering, and short connection paths, similar to other larger social networks. 
BlueSky introduced feeds—algorithmic content recommenders created for and by users. We analyze all feeds and find that while a large number of custom feeds have been created, users' uptake of them appears to be limited. 
We analyze the hyperlinks shared by BlueSky's users and find no evidence of polarization in terms of the political leaning of the news sources they share. They share predominantly left-center news sources and little to no links associated with questionable news sources. 
In contrast to the homogeneous political ideology, we find significant issues-based divergence by studying opinions related to the Israel-Palestine conflict. Two clear homophilic clusters emerge: Pro-Palestinian voices outnumber pro-Israeli users, and the proportion has increased. We conclude by claiming that Bluesky—for all its novel features—is very similar in its network structure to existing and larger social media sites and provides unprecedented research opportunities for social scientists, network scientists, and political scientists alike.
\end{abstract}

\section*{Introduction} 
Bluesky is a novel and decentralized social media site that opened up in an invite-only beta release in February 2023. The network is a microblogging site, explicitly describing itself as ``a Twitter-style social app'' \cite{kleppmann2024bluesky}. In 2019, Bluesky originated as the ``Bluesky initiative'' and was announced by the then CEO of Twitter, Jack Dorsey \cite{dorsey2019}. As a separate entity, the takeover of Twitter (now X) by Elon Musk did not affect Bluesky's operation. The website has grown to 5,7 million users \cite{bsky_stats2023}. In early February of 2024, Bluesky opened the website to users without an invite. 

While Bluesky is modeled after Twitter, it sets out to solve the ``thorniest problems of social media'' such as misinformation, harassment, and hate speech by implementing decentralization and leveraging a marketplace approach to
these problems\cite{kleppmann2024bluesky}. 
Notably, Bluesky allows anybody to create custom moderation and algorithmic curation services, and users are free to subscribe to the ones they prefer.
Decentralization means that the protocol or platform draws upon ``multiple interoperable providers for every part of the system''. In practice, it means that several competing clients for the platform such as Graysky \cite{graysky2023} and deck.blue \cite{deckblue2023}, and the official Bluesky app are available to each user. Additionally, users can self-host their data on Personal Data Servers (PDS), which store user data and allow other participants of Bluesky to query their data \cite{kleppmann2024bluesky}. Bluesky acknowledges that most users will sign up on a shared PDS run by a professional hosting provider. This provider, however, does not need to be Bluesky; it can be run by anyone. Bluesky's decentralized design allows broad data access and the range of choices given to users is unprecedented for a large social media site \cite{failla2024m}.

In this study, we examine the complete Bluesky network. We investigate users' activity on Bluesky over time and provide the first insights into what drove user sign-ups during the website's rise. 
Complete and longitudinal studies of a social media platform's evolution are rare as they require extensive time-series data.

Consequently, there have been relatively few studies that capture the full developmental trajectory of a social media network from its inception. For example, previous studies have looked at the right-wing social media site Parler and the decentralized platform Mastodon \cite{zignani2018follow} \cite{aliapoulios2021early}.
Here, we characterize the topology of the network of Bluesky throughout the observation period.  We describe the network based on various degree distributions, its clustering, density, and connectivity. 

In a backlash against ``opaque content recommendation systems'' \cite{kleppmann2024bluesky}, multiple decentralized social networks, such as Mastodon, implemented non-algorithmic, reverse chronological feeds \cite{zignani2018follow}. Bluesky sets out to give users more agency over their own user experience. In practice, users have more choices in moderation and can design and subscribe to diverse content recommendation algorithms. Bluesky enables users to generate feeds letting them ``choose their algorithms'' in an effort to aid users discover content from other users they do not know and to gain exposure to specific content posted. Bluesky's innovative feed feature has been used to create over 39 thousand feeds by a subset of highly active users. Feeds showcases a wide array of algorithmic choices, ranging from simple regex-based filters to professionally curated content streams. Popular feeds include ``Discover'' which promises to show ``Trending content from your personal network'' \cite{bsky_whats_hot}, ``Mutuals'' showing posts from followers of the user, but also topic-specific collections such as ``Science'' and ``Art''. Notwithstanding the breadth of creative feeds, the feature's overall adoption appears limited relative to Bluesky's total user base. Engagement with feeds follows a familiar pattern in social media: a heavy-tailed distribution where a small number of feeds and highly active users dominate the landscape. This aspect of the platform provides an opportunity to investigate how users engage with and potentially influence content curation mechanisms, a topic of growing importance in the study of social media dynamics.

Over the last decade, social media has become more fragmented with an increasing number of smaller or fringe platforms, serving a cohesive group of users \cite{kubin2021role, stocking2022role}.  Recent years have seen the proliferation of smaller, niche social media platforms catering to specific user groups or ideologies. These platforms often exhibit high levels of homogeneity in terms of user demographics and political leanings, contrary to platforms such as Twitter and Facebook that show a polarization across political ideologies\cite{flamino_political_2023,gonzalez-bailon_asymmetric_2023}. For instance, Gab and Parler have been found to attract predominantly conservative users  \cite{gerard2023truth, sharevski2022gettr, aliapoulios2021early, than2021welcome}. While this homogeneity can create strong communities, it also raises concerns about echo chambers and the potential for increased polarization. However, an open question is whether, within these more homogeneous platforms, specific issues or topics may still show a polarization of opinions. This study investigates whether Bluesky, as a new and growing platform, exhibits similar patterns of homogeneity observed in other small platforms\cite{gerard2023truth, sharevski2022gettr, aliapoulios2021early, than2021welcome} or if it manages to attract a more diverse user base. We examine both broad political leanings and specific issue-based discussions to provide an understanding of the platform's user composition and potential for diverse discourse. We look at the spread of misinformation, finding that Bluesky users disseminate little to no information associated with news sources associated with conspiracy theories, propaganda, or fake news. Of all posts containing domains posted to Bluesky only 0.14\% are classified as questionable. This small number of posts was authored by 3,704 unique users, making the spread of news marked as questionable almost nonexistent on Bluesky. Next, we investigate the political ideology on a left-right spectrum and show that Bluesky is almost homogeneously left-center biased, with very few users having a right-of-center ideology. Lastly, we show that conversations surrounding the Israel-Palestine conflict are highly polarized, indicating that political homogeneity does not necessarily dictate consensus on specific issues.

\section*{Results}
\subsection*{Activity on Bluesky} \label{sec:actonbl}
\begin{figure}[!ht]
    \centering
    \includegraphics[width=\columnwidth]{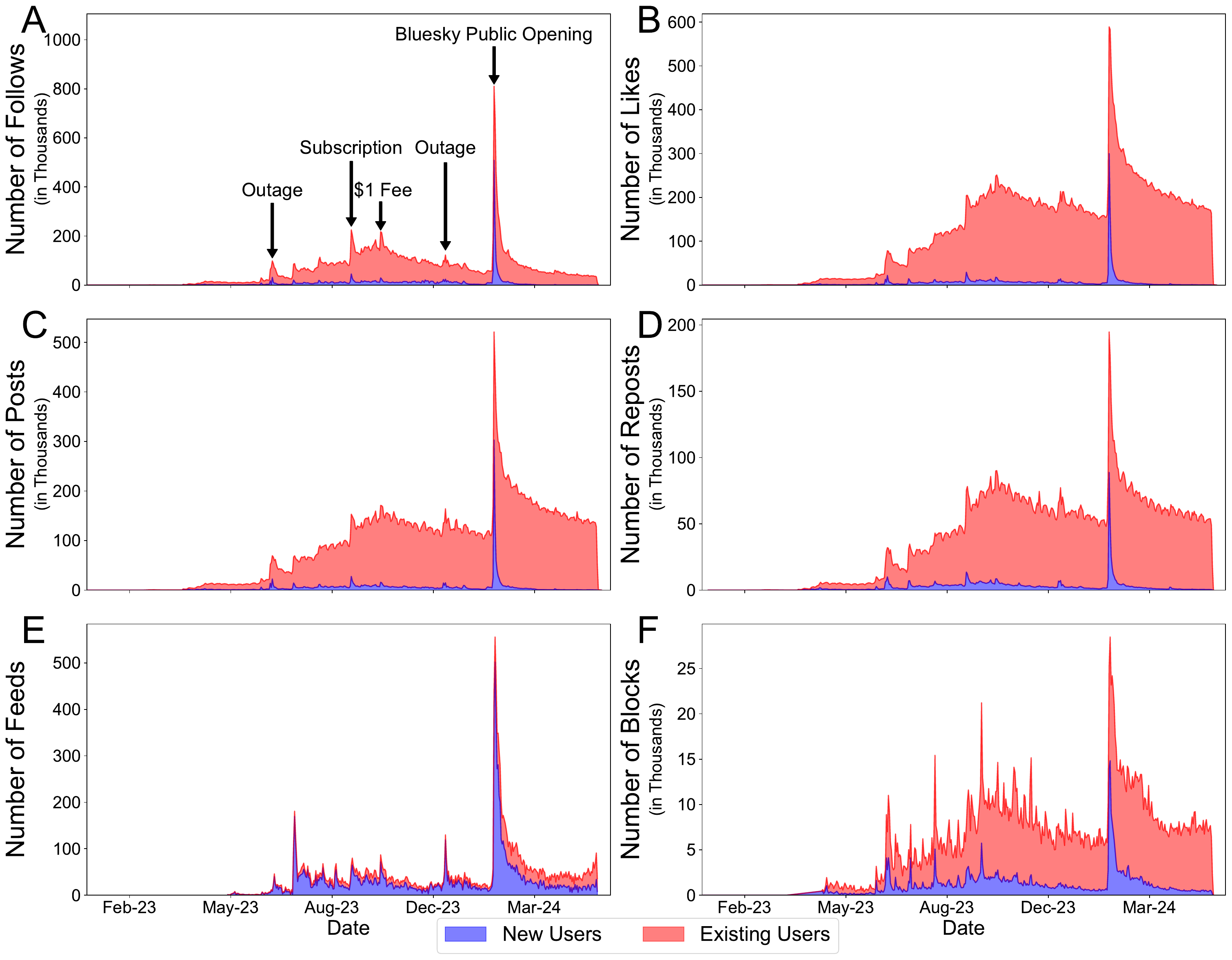}
    \caption{User activity on the BlueSky social media platform from February 2023 to May 20, 2024. Each panel details the number of new and existing active users, ranging from follows (\textbf{A}), likes (\textbf{B}), posts (\textbf{C}), reposts (\textbf{D}), feed generation (\textbf{E}), to blocks (\textbf{F}), showing the number of unique users engaging through these actions. The term 'New Users' refers to individuals interacting for the first time with the platform through the respective activity measure. Blue areas denote new users, while red areas show the number of existing users engaging in activity.}
    \label{fig:user_activity_bluesky}
\end{figure}

Figure \ref{fig:user_activity_bluesky} presents the daily number of active users according to six engagement metrics. Dates with the highest numbers of new users over the year 2023 were substantially driven by activity and news about X (formerly Twitter). While further research leveraging qualitative surveys is necessary to establish the exact reasons for users switching, the number of sign-ups significantly correlates with news about Twitter \cite{jeong2024user}.

We look at days with a proportionally high number of new sign-ups by calculating the ratio of new to existing users on the platform over the course of 2023. On September 19, the day with the highest ratio of new to old users, X announced that all users might be charged a fee to use the website (``Elon Musk says Twitter, now X, could charge all users subscription fees'' \cite{guardian2023subscription}, ``Elon Musk: Social media platform X, formerly Twitter, could go behind paywall'' \cite{bbc2023subscription}). September 19 and 20 saw the first and fourth-highest numbers of new active users to existing users on the platform, respectively. The day with the second most sign-ups, relative to the size of the active userbase, was July 3, 2023. On this day, X experienced global outages as a bug caused users to receive rate-limit errors, preventing them from viewing an unlimited number of posts (``Twitter rate-limits itself into a weekend of chaos'' \cite{register2023ratelimit}, ``Twitter’s Troubles Are Perfectly Timed for Meta'' \cite{bloomberg2024troubles}). October 18 and 19 experienced the third and fifth-highest ratios of new users to existing users engaging with Bluesky. On October 18, Twitter announced a \$1 fee for new users in New Zealand and the Philippines (``X, formerly Twitter, rolls out US \$1 annual fee for new users in New Zealand and the Philippines'' \cite{guardian2023fee}, ''Starting today, we're testing a new program (Not A Bot) in New Zealand and the Philippines. New, unverified accounts will be required to sign up for a \$1 annual subscription to be able to post \& interact with other posts.'' \cite{twitter2023support}). Lastly, on December 21, Twitter experienced another global outage, leading to another surge of sign-ups on Bluesky (``X, formerly Twitter, sees massive outage as tens of thousands report issues'' \cite{euronews2023outage}, ``Is X/Twitter down? Users report problems accessing feeds in multiple countries'' \cite{independent2023outage}). In 2024, the week following Bluesky's opening to the public on the 7th of February 2024 surpassed previous records of new active users compared to the existing user-base. We see very similar patterns across all usage metrics. Following the inception of Bluesky activity slowly grew up to a peak in mid 2023 when activity on the platform slowly decreased. The public opening of Bluesky lead to an unprecedented peak with activity quickly decaying to levels of activity similar to the summer of 2023. Activity on Bluesky now seems to have stabilised for now.

\subsection*{Evolution of the Bluesky Network}
The data gathered via the Bluesky API represents a temporal network of the entire interaction graph of the social media platform. This allows us to analyze changes in the network topology over time.

Social media sites such as Bluesky are often described as a singular ``network'' connecting users to each other. However, users of social media sites form relationships and interact with users through a variety of different mechanisms - all capturing different relationships that are not necessarily ontologically equivalent \cite{dickison2016multilayer}. Magnani and Rossi \cite{magnani2011ml} find large differences in the centrality of users on social media depending on the interaction layer they investigate. We, therefore, describe the topological structure of Bluesky based on four distinct layers: Followership, Replies, Reposts, and Likes. For a description of the interactions underlying the individual layers, please refer to the Materials and Methods section. 

\begin{figure}[!ht]
    \centering
    \includegraphics[width=\columnwidth]{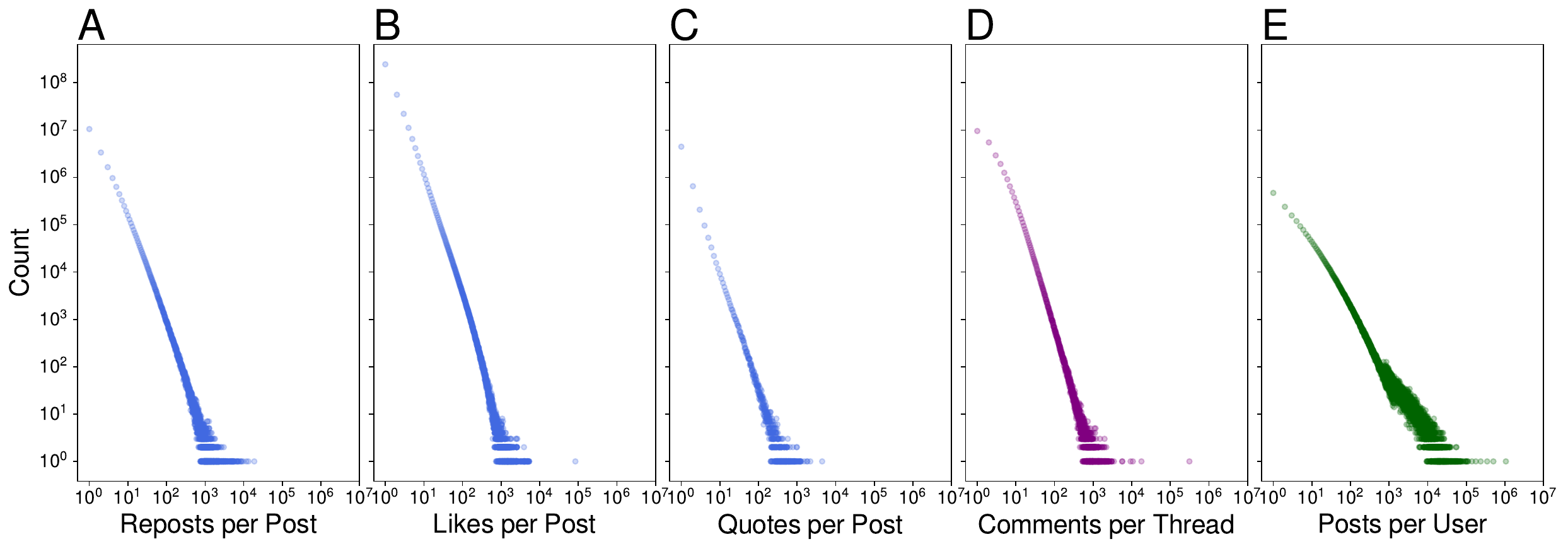} 
    \caption{Social media interaction metrics with posts (log-log scale). Each panel plots a specific metric against its frequency to analyze patterns of user engagement and content spread. The X-axis represents the specific metric, and the Y-axis shows the frequency of occurrences for each metric value. (\textbf{A}) Reposts per Post. (\textbf{B}) Likes per Post. (\textbf{C}) Quotes per Post. (\textbf{D}) Comments per Thread. (\textbf{E}) Posts per User. All plots use logarithmic scales.}
    \label{fig:postdistribution}
\end{figure}

\begin{figure}[!ht]
    \centering
    \includegraphics[width=\columnwidth]{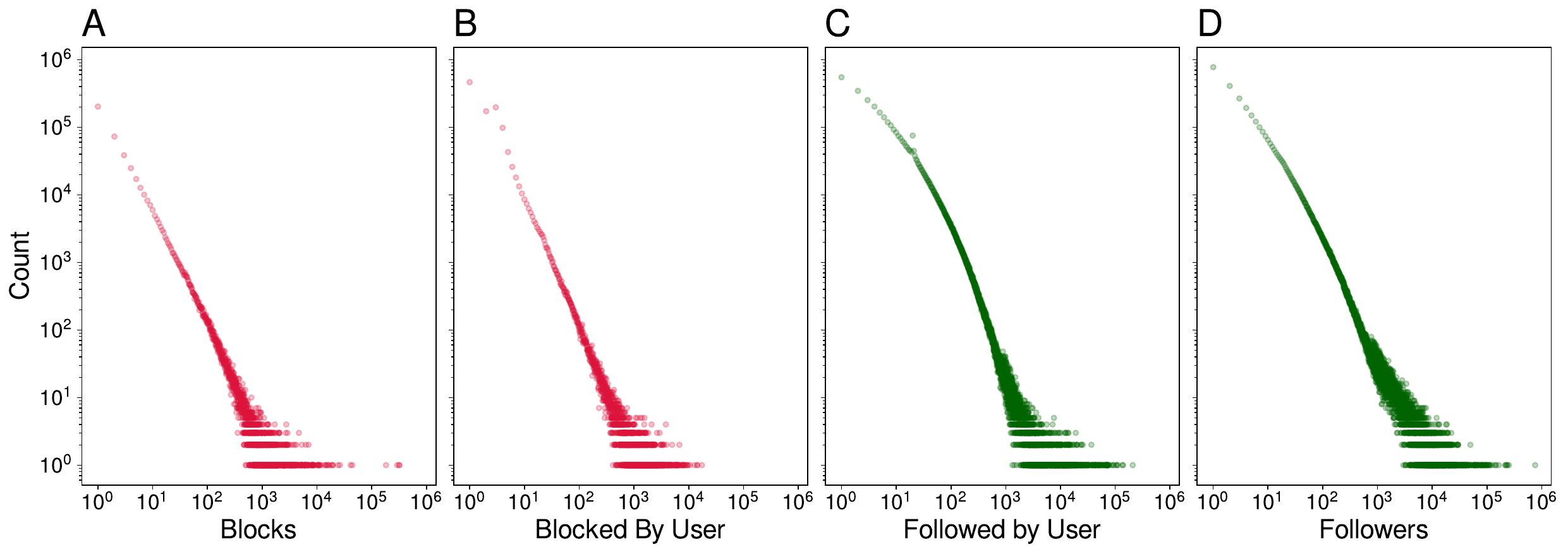}
    \caption{Social media interaction metrics by Users (log-log scale). Each panel represents a specific interaction metric plotted against its occurrence frequency to analyze patterns of user interactions. The X-axis denotes the metric in question, while the Y-axis shows the frequency of occurrences for each metric value. (\textbf{A}) Reposts per Post.(\textbf{B}). Likes per Post. (\textbf{C}) Comments per Thread. (\textbf{D}) Posts per User.}
    \label{fig:userdistribution}
\end{figure}

Figures \ref{fig:postdistribution} and \ref{fig:userdistribution} show the distribution of engagement metrics per post and engagement metrics per user. All distributions exhibit a power-law distribution with a large number of users and posts receiving or authoring few interactions, and a small number of entities being responsible for the vast majority of interactions. 

For all distributions we report the mean $\mu$, standard deviation $\sigma$, skewness $\gamma$, kurtosis $\beta$, minimum $m$, maximum $M$, the exponent of the power-law distribution $\alpha$ and the ratio of mean to maximum  $\frac{\mu}{M}$ in table \ref{tab:DistributionValues}. 

\begin{table}[!ht]

    \centering
        \begin{tabular}{lrrrrrrrr}
            \toprule
            Distribution & $\mu$ & $\sigma$ & $\gamma$ & $\beta$ & m & M & $\frac{\mu}{M} * 10^3$ & $\alpha$ \\
            \midrule
\# Quotes per Post & 1.682 & 6.003 & 149.259 & 67646.663 & 1 & 4455 & 0.377 & 5.441 \\
\# Likes per Post & 2.148 & 9.521 & 5355.320 & 53468722.557 & 1 & 102123 & 0.021 & 3.702 \\
\# Reposts per Post & 4.140 & 20.348 & 141.972 & 70043.802 & 1 & 18765 & 0.221 & 2.503 \\
Thread Length & 4.185 & 62.836 & 4780.687 & 23736718.688 & 1 & 312092 & 0.013 & 2.145 \\
\# People Blocked By & 8.155 & 81.228 & 70.535 & 9086.269 & 1 & 18234 & 0.447 & 2.122 \\
\# People Blocked & 19.494 & 838.448 & 340.712 & 123106.559 & 1 & 324147 & 0.060 & 1.900 \\
\# People Followed by User & 38.952 & 343.076 & 203.031 & 71572.734 & 1 & 208288 & 0.187 & 1.441 \\
\# Followers & 43.231 & 668.036 & 516.367 & 498518.321 & 1 & 755584 & 0.057 & 1.527 \\
\# Posts per User & 89.601 & 1061.988 & 501.547 & 450864.562 & 1 & 1059206 & 0.085 & 1.458 \\
            \bottomrule
        \end{tabular}
    \caption{User activity distributions are indicative of heavy-tailed behavior. Included metrics are the mean $\mu$, standard deviation $\sigma$, skewness $\gamma$, kurtosis $\beta$, minimum $m$, maximum $M$, and the ratio of mean to maximum  $\frac{\mu}{M}$ from the distributions depicted in Fig. \ref{fig:postdistribution} and \ref{fig:userdistribution}. The exponent of the power law distribution is denoted by $\alpha$ . The notably high values of $\gamma$ and $\beta$ indicate a pronounced right-skewed, heavy-tailed nature across all distributions. Furthermore, the exceptionally low $\frac{\mu}{M}$ values further confirm the extensive tail behavior characteristic of these distributions. ''\#'' should be read as Number of.}
    \label{tab:DistributionValues}

\end{table}

\begin{figure}[!ht]
\centering
\includegraphics[width=\columnwidth]{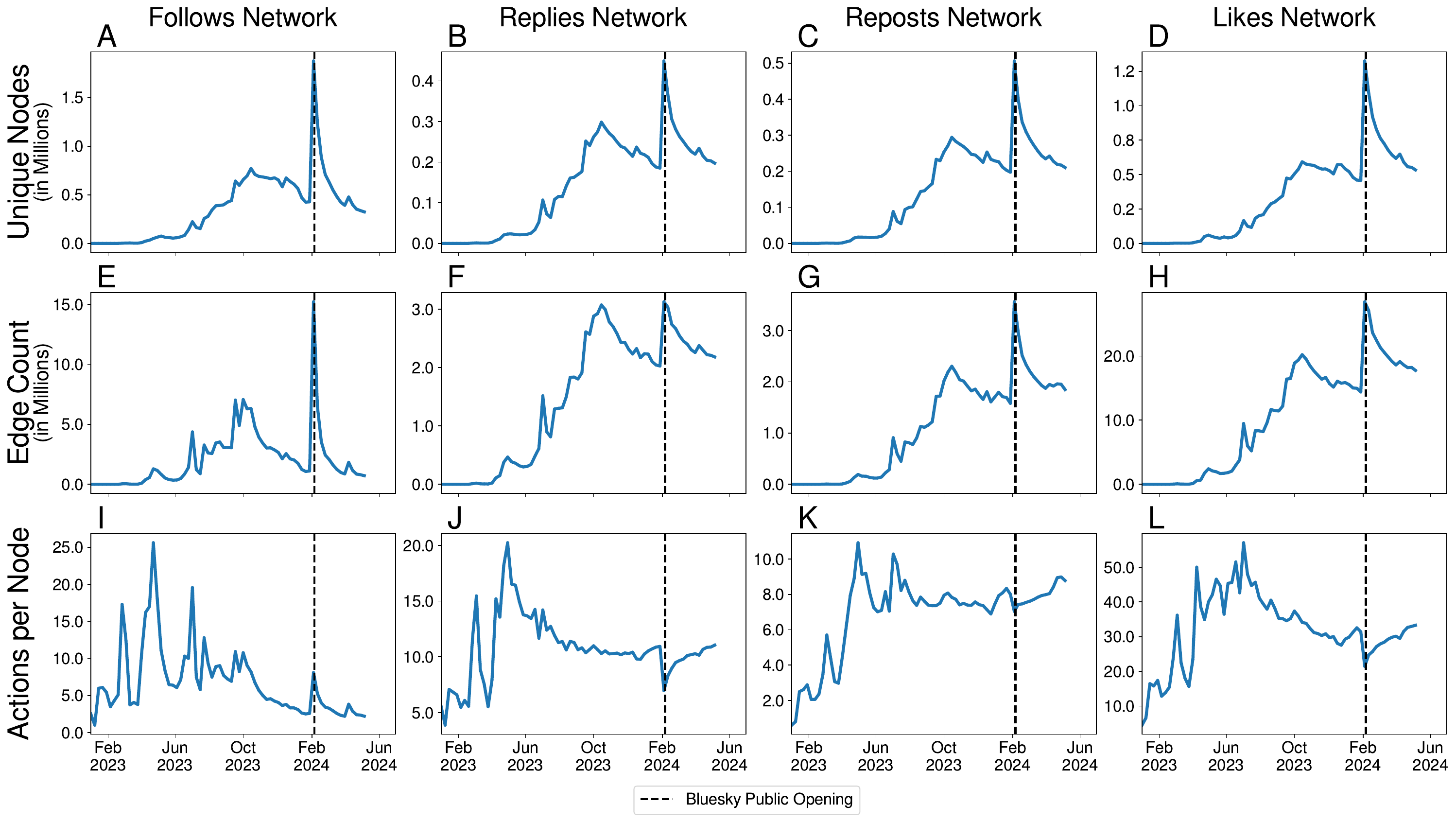}
\caption{Metrics capturing changes in the network structure from 2023 to May 2024. These metrics are computed across four networks. (\textbf{A}-\textbf{D}) Count of unique nodes (in millions) active per week for each network. (\textbf{E} - \textbf{H}) Number of unique Edges in the network per Week. (\textbf{I} - \textbf{L})  Ratio of edges to unique edges, capturing the activity of nodes in each week. The black dashed line in each graph denotes the date of the public opening of Bluesky.}
\label{fig:cc-evolution_act}
\end{figure}

Figure \ref{fig:cc-evolution_act} shows three key metrics that chart changes in the network structure from 2023 to May 2024, computed across four distinct networks. Figures \ref{fig:cc-evolution_act}\textbf{A} to \ref{fig:cc-evolution_act}\textbf{D} illustrate the weekly count of unique, active users for each network. A clear growth trend in user engagement is observed from February 2023 until September 2023, with peak activity observed at different magnitudes across networks—700,000 users in followership, 300,000 in both replying and reposting, and 600,000 in the Likes network. After these peaks, there is a notable decline in activity, which reverses in February 2024 following Bluesky’s public launch, allowing unrestricted user access. This results in record-high weekly activities across the networks: 2 million users in followership, 450,000 in replying, 550,000 in reposting, and 1.3 million in Likes. Since the opening of the platform, the number of follows, comments, reposts and likes on the platform has been slowly decreasing to levels last seen before the opening of Bluesky. The steepest drop in activity is seen in the Follows network. This is likely driven by the initial actions of new users who follow suggested profiles without further significant engagement.

The figures \ref{fig:cc-evolution_act}\textbf{E} to \ref{fig:cc-evolution_act}\textbf{H} depict the weekly number of interactions within each network, showing trends similar to those of user engagement. The growth in interactions peaks in September 2023 across all networks—6 million in followership, 3.2 million in replying, 2.2 million in reposting, and 21 million in Likes—before decreasing and then surging to new highs in February 2024. The followership interactions show the highest variability, mirroring the bursty nature of sign-ups on the platform.

Lastly, figures \ref{fig:cc-evolution_act}\textbf{I} to \ref{fig:cc-evolution_act}\textbf{L} focus on the average interactions per unique user within each network. The metrics climb until mid-2023, reaching their zenith in April for Followership with 26 interactions, May for replying with 21 interactions, April for Reposting with 12 interactions, and July for Likes with 60 interactions. A gradual decrease follows until 2024. Upon Bluesky’s public opening, there's a noticeable dip in average interactions for Likes, Reposts, and replying, indicating lower activity levels among newer users compared to the earlier, invitation-only cohort. Conversely, the average interactions in the Followership network increase, suggesting that the newly joined users are relatively more engaged in following activities than in other forms of interaction. After the influx of new users the average activity per active user has steadily increased for all but the following network. 

\begin{figure}[!ht]
    \centering
    \includegraphics[width=\columnwidth]{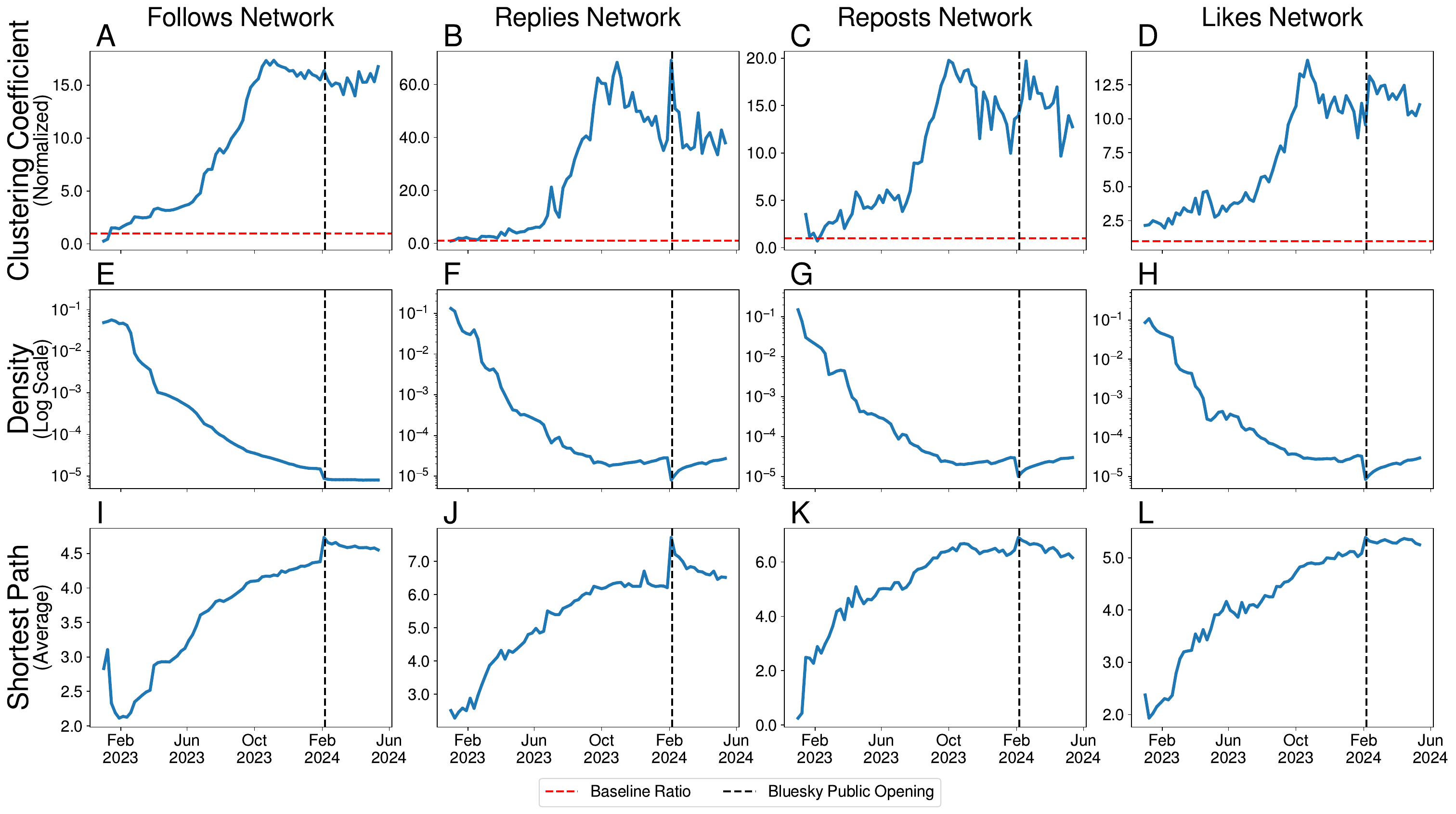}
    \caption{Metrics capturing changes in the network structure from 2023 to May 2024. Clustering-coefficient, density, and average shortest path are computed for four networks. Replies, Reposts, and Likes capture non-persistent interactions, thus all metrics are calculated individually for each week's edges. The followership network is persistent. (\textbf{A}-\textbf{D}) Normalized average clustering coefficient. The dashed red line represents an equal value for the random and original graph. (\textbf{E}-\textbf{H}) Density of the networks.
    (\textbf{I} - \textbf{L}) Average shortest path length for all networks.
    The black dashed line in each graph denotes the date of the public opening of Bluesky.}
    \label{fig:ccevolutionmeas}
\end{figure}

Figure \ref{fig:ccevolutionmeas} shows three measures capturing changes in the structure of the network over the observation period. Figure \ref{fig:ccevolutionmeas}\textbf{A} to figure \ref{fig:ccevolutionmeas}\textbf{D} focus on the normalized average clustering coefficient for each network, a measure that is adjusted by comparing it to a randomized graph with the same degree sequence. This comparison is visualized where the dashed red line indicates parity between the real and randomized networks. The consistent observation that the normalized clustering coefficient remains above one suggests that the network structure is more cliquish than random models would predict. Similarly to figure \ref{fig:ccevolutionmeas}\textbf{A} to figure \ref{fig:ccevolutionmeas}\textbf{H}, the normalized clustering coefficient increases gradually until September of 2023, where activity on Bluesky was locally maximal. The magnitude of peaks varies with the Followership network reaching a coefficient of 10, while Reposts and Likes peak at coefficients of 10 and 16, respectively. The replies network, however, achieves a significantly higher peak of 200, indicating exceptionally dense clustering. Following these peaks, all coefficients trend downwards until February 2024 due to new user influx. While the clustering coefficient remains volatile for the non-persistent interaction, it seems to have somewhat stabilized. 

Figure \ref{fig:ccevolutionmeas}\textbf{E} to \ref{fig:ccevolutionmeas}\textbf{H} present the density of these networks over time, which reflects the proportion of actual connections relative to the maximum possible. All density values are plotted on a logarithmic scale to highlight trends more clearly. Despite the non-persistent nature of the three networks, all exhibit a consistent, sub-linear decline in density over time. This downward trend is accentuated in February 2024, when network density sharply decreases across all networks due to the sudden increase in the user base following Bluesky's public opening.

Figure \ref{fig:ccevolutionmeas}\textbf{I} to \ref{fig:ccevolutionmeas}\textbf{L} show the average shortest path in all networks over time. For all networks we observe a slow and sub-linear increase in the average distance until February 2024. When Bluesky opened up to the public, the average distance sharply increases, to varying extents, across all networks. This can be attributed to new users which are only loosely connected to the network. After these users connected to the network, we observe a slow decline in the average distance for the replies and reposts network, and stagnation in the followership and likes network. Importantly, for all networks the average shortest path remains very low showing the connectivity, efficiency, and small-worldness of the Bluesky network.

\subsection*{Feeds}
Unlike traditional social media platforms, Bluesky introduces a novel ``marketplace of algorithms''\cite{kleppmann2024bluesky} through its feed feature, enabling users to design, implement, and distribute their own content curation mechanisms, ranging from simple keyword matching to complex machine learning models \cite{jeong2024bluetempnet}. User-generated feeds are a core functionality of Bluesky and an alternative to ``opaque content recommender systems'' used by larger platforms. Kleppmann et al. \cite{kleppmann2024bluesky} cite feeds created by users based on regex matching and machine learning algorithms. Other feeds leverage the network structure of Bluesky and surface content from users' followers. The default algorithm for Bluesky users is a non-algorithmic reverse-chronological feed of their connections.

In total, 39,639 feeds have been created by 18,352 active users showcasing the breadth of content curation algorithms available to users on Bluesky and the broad usage of this novel feature. Users can bookmark a feed \cite{failla2024m}, which pins the feed on their home screen. While bookmarks are private on Bluesky, it is public knowledge whether users a ``liked'' a feed. In our dataset, 139,033 Users have used this feature and liked feeds 295,902 times. 

\begin{table}[!ht]
    \centering
    \begin{tabular}{llr}
    \toprule
     Displayname & Description & Number of Likes \\
    \midrule
     For You & Learns what you like & 16,132 \\
     OnlyPosts & Posts from people you follow without ... & 5,137 \\
     Science & The Science Feed. A curated feed from ... & 4,972 \\
    Adult Content & Formerly ``Suggestive'' All (nonviolent) ... & 4,180 \\
     Art & Images posted by artists on Bluesky... & 3,268 \\
     New & Posts by furries across Bluesky... & 3,221 \\
     Mutuals & Posts from users who are following you... & 3,103 \\
     Home+ & Its the Home feed Blue Sky was missing... & 3,050 \\
     Art & Posts by furries with \#furryart... & 2,938 \\
     Blacksky & Amplifying the voices of any and all... & 2,778 \\
    \bottomrule
    \end{tabular}
    \caption{Top Feeds by number of likes on Bluesky.}
    \label{tab:TopFeeds}
\end{table}

The most liked feed \emph{For You} has been renamed \emph{Discover} and promises to show ''Trending content from your personal network'' \cite{bsky_whats_hot}. Other popular feeds include ''Science'', which is a feed curated by ''Bluesky professional scientists'' \cite{bsky_for_science}.
Other feeds such as the ''Hospitality \& Tourism'' or ''Paleo Sky'' use regex patterns to match posts (\emph{tourism, skift} and \emph{Paleontology|Archaeology|\#PaleoSky} respectively). 

\begin{figure}[!ht]
    \centering
    \includegraphics[width=\columnwidth]{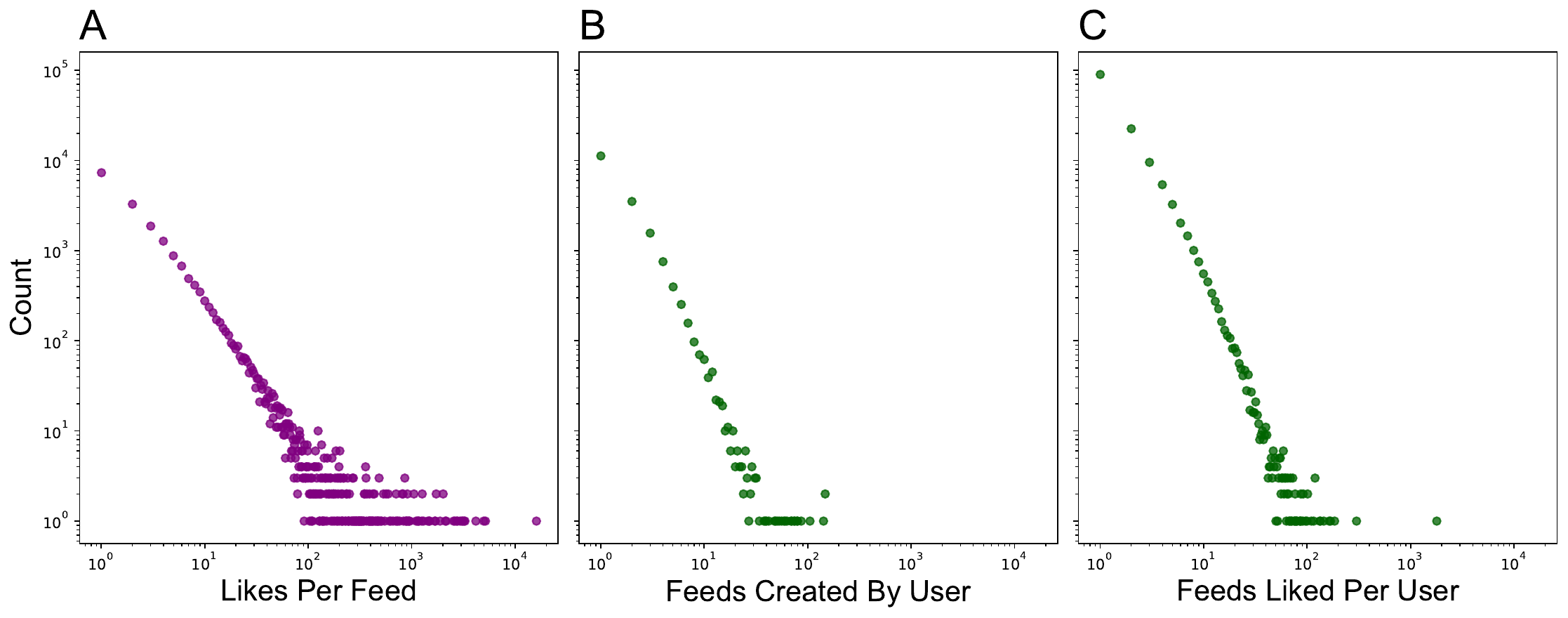}
    \caption{Log-log plots displaying distribution of feed related statistics. (\textbf{A}) Number of likes received per feed. (\textbf{B}) Number of feeds created per user. (\textbf{C}) Number of feeds liked per user.}
    \label{fig:feeddistribution}
\end{figure}

Figure \ref{fig:feeddistribution} shows the distribution of the number of likes per feed, the number of feeds created per user, and the number of feeds liked per user. 
Table \ref{tab:TopFeeds} shows the most liked feeds in our dataset. We again see heavy-tailed distributions with the most active participants liking and creating exponentially more content than the median user.

 Table \ref{tab:feeddistr} summarizes the descriptive statistics of the distributions. On average, feeds only attracted two likes - showing that most feeds receive little to no engagement. Conversely, users who actively liked feeds liked, on average, over fourteen distinct feeds. We also see that a large number of feeds was created by a small minority of highly active users. This indicates that users who take advantage of this feature express their preferences for algorithmic choices broadly, which could help researchers study algorithmic choices. However, in total, only 139 thousand out of 5 million users liked at least one feed.

\begin{table}[!ht]
    \centering
    \begin{tabular}{lrrrrrrrr}
    \toprule
    Distribution & $\mu$ & $\sigma$ & $\gamma$ & $\beta$ & m & M & $\frac{\mu}{M} * 10^3$ & $\alpha$ \\
    \midrule
    \# Likes per Feed & 2.128 & 5.932 & 203.296 & 60,656.913 & 1 & 1,799 & 1.183 & 3.413 \\
    \# Feeds Created per Person & 2.161 & 13.931 & 122.996 & 16,082.634 & 1 & 1,828 & 1.182 & 3.350 \\
    \# Number of Feeds Liked Per User & 14.783 & 156.641 & 62.968 & 5,800.619 & 1 & 16,132 & 0.916 & 1.877 \\
    \bottomrule
    \end{tabular}
    \caption{Activity related to Feeds on Bluesky. Included metrics are the mean $\mu$, standard deviation $\sigma$, skewness $\gamma$, kurtosis $\beta$, $m$, maximum $M$, and the ratio of mean to maximum  $\frac{\mu}{M}$ from the distributions depicted in Fig. \ref{fig:postdistribution} and \ref{fig:userdistribution}. $\alpha$ indicates the exponent of the power law distribution. The notably high values of $\gamma$ and $\beta$ indicate a pronounced right-skewed, heavy-tailed nature across all distributions. Furthermore, the exceptionally low $\frac{\mu}{M}$ values further confirm the extensive tail behaviour characteristic of these distributions.}
    \label{tab:feeddistr}
\end{table}

\subsection*{Political Leaning \& Polarization of BlueSky}
Small and novel social media platforms have oftentimes been characterized by little diversity in political and ideological viewpoints. Truth Social - launched in February of 2022, leans strongly conservative - and was created as an ``alternative social media platform'' targeting Republican social media users \cite{gerard2023truth}. Voat.co, a small Reddit-esque social network, grew after Reddit banned thousands of subreddits, attracting banned extreme communities \cite{mekacher2022can}. Similarly, Gab was founded to attract alt-right users \cite{sharevski2022gettr} and Parler became the home of ``disaffected right-wing social media users'' \cite{aliapoulios2021early}. While Mastodon has elements of techno-libertarian leaning \cite{nicholson2023mastodon} and took active measures to distance themselves from right-wing users \cite{gehl2023digital}, there exists to our knowledge no study examining the political leanings of users of the platform. While the literature contains ample examples of small social media platforms launched because of content moderation on Twitter, which were perceived to be disproportionally targeting conservative viewpoints \cite{sharevski2022gettr}, there is, to our knowledge, no empirical investigation into a predominantly left-leaning small platform. 

Twitter, now X, is a platform that has a substantial user base with diverse political ideologies, ranging from far-right to far-left, although the majority of the user base has been characterized as left-leaning/center \cite{wojcik2019sizing}. As shown in section \emph{Activity on Bluesky}, sign-ups to Bluesky have been driven by activity on Twitter and its new leadership under Elon Musk. Since the purchase of Twitter by Elon Musk and its subsequent rebranding as X (the everything app), several newspapers and academics have reported that the user base, moderation philosophy, and goals of the platform have shifted towards a more right-leaning approach \cite{pew2023twitter, economist2023twitter, guardian2023twitter}. The perception of a shift towards the right on Twitter and the correlation of news about Twitter with sign-ups to Bluesky lead us to expect Bluesky to be predominantly left-wing, consisting of users who left the platform in search of a new social media site that is closer to their ideology. However, issue and platform polarization are hard to predict and strongly influenced by first movers and path dependencies \cite{macy2019opinion}. We investigate the political leaning of Bluesky by extracting the domain of all links shared on Bluesky over the entire observation period. Table \ref{tab:source_domains} lists the most shared Non-Political, Political, and ``Questionable-Source'' domains based on ratings by Media Bias Fact Check (MBFC) \cite{mediabiasfactcheck}. We classify a website as ``Political'' if its domain has an associated MBFC rating. We also report overall domain counts. Lastly, we show all ``Questionable-Source'' websites, filtered to include only those categorized by MBFC as either spreading fake news, conspiracies, or propaganda.

\begin{table}[!ht]

\centering
\begin{tabular}{@{\extracolsep{5pt}} cc cc cc}
\toprule
\multicolumn{2}{c}{Overall} & \multicolumn{2}{c}{Political} & \multicolumn{2}{c}{Questionable Sources}\\
\cmidrule(r){1-2}\cmidrule(l){3-4}\cmidrule(l){5-6}
Source Domain & Count & Source Domain & Count & Source Domain & Count\\
\cmidrule(r){1-1}\cmidrule(lr){2-2}\cmidrule(lr){3-3}\cmidrule(l){4-4} \cmidrule(lr){5-5}\cmidrule(l){6-6}
youtube.com & 1,665,744 & theguardian.com & 124,123 & dailymail.co.uk & 2,501 \\
spotify.com & 237,786 & nytimes.com & 82,963 & mondoweiss.net & 2,249 \\
tenor.com & 185,865 & washingtonpost.com & 33,388 & foxnews.com & 870 \\
twitch.tv & 185,268 & tagesschau.de & 29,454 & newsbreak.com & 430 \\
theguardian.com & 124,123 & bbc.com & 27,973 & indiatimes.com & 401 \\
substack.com & 86,116 & cnn.com & 26,838 & thenationalnews.com & 318 \\
instagram.com & 83,077 & spiegel.de & 23,757 & moveon.org & 279 \\
nytimes.com & 82,963 & apnews.com & 23,305 & almayadeen.net & 270 \\
twitter.com & 77,440 & reuters.com & 23,230 & theepochtimes.com & 262 \\
openstreetmap.org & 68,503 & theverge.com & 21,578 & presstv.ir & 244 \\
\cmidrule(r){1-2}\cmidrule(l){3-4}\cmidrule(l){5-6}
Total: & 8,409,741 & Total: & 1,582,455 & Total: & 11,984 \\
\% of Total: & (100\%) & \% of Total:  & (18.81\%) & \% of Total: & (0.14\%) \\
\bottomrule
\end{tabular}
\caption{Top ten domains by frequency in the dataset, comparing the overall occurrences to those classified under political and questionable sources (restricted to fake news, propaganda, or conspiracy) categories, respectively. To filter automated accounts, we exclude posts from accounts with more than 10,000 posted URLs.}
\label{tab:source_domains}

\end{table}

The most frequent non-political domains mostly relate to other social media websites. To ensure that we correctly mapped all links to domains, we expanded all links associated with a list of link-shorteners \cite{sambokai_ShortURL_Services_List}. YouTube, the domain with the highest number of shares, was linked to a total of 1.66 million times. The second most shared domain is Spotify.com. Other frequently shared social media domains include Twitch (185,268), Twitter (77,440), Instagram (83,077), and Substack (86,116) links.  Tenor is a Gif sharing website. Interestingly, two political domains are among the most shared domains. \emph{The Guardian} and the \emph{New York Times} are classified by MBFC as ``left-center''. In total, 408,133 unique domains were posted 8,409 million links to Bluesky.

With the exception of \emph{Tagesschau} and \emph{Reuters}, all political outlets in the top ten political outlet columns are classified by MBFC as ``left-center''. The two exceptions are both classified as ``center'' (or least biased). Prior to analyzing the overall distribution of all political domains in the dataset, this already indicates the bias of the platform. In total, we observed 1,582,455 occurrences of political domains being spread on Bluesky, making up 18.81\% of all posts that include links in the dataset.

Compared to the spread of political domains, there is very little information stemming from websites classified as questionable sources, being spread. The top questionable source domain is``dailymail.co.uk'' with only 2,501 occurrences in the entire dataset. Less than a percentage of posts including links on Bluesky contain links to domains classified as spreading fake news, conspiracies, or propaganda. The top 150 users spreading questionable news-sources make up 50\% of the total spread on the platform. In total only 3,704 users on Bluesky have ever posted a domain associated with fake news, conspiracies, or propaganda.

\begin{figure}[!ht]
    \centering
    \includegraphics[width=\columnwidth]{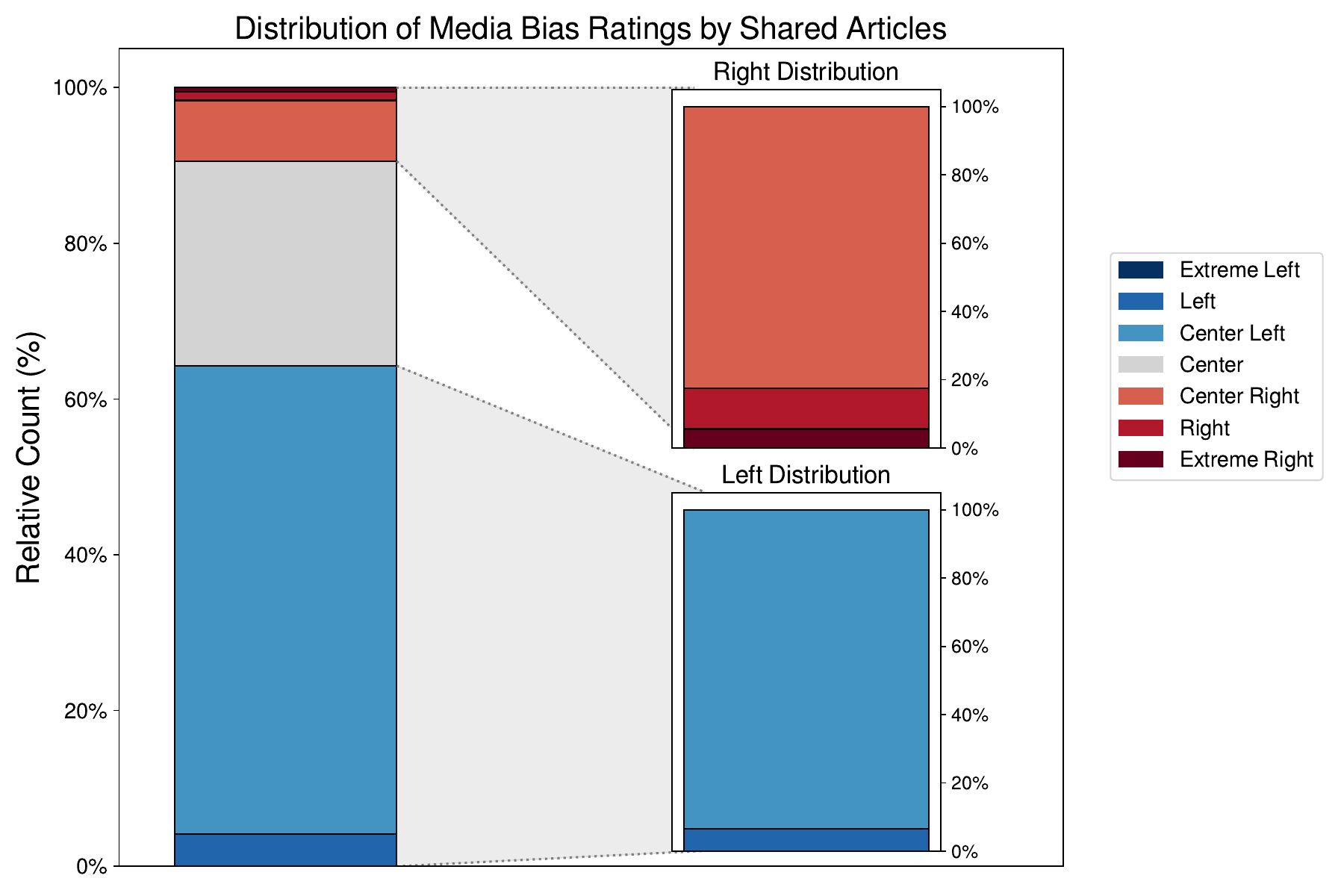}
    \caption{Distribution of political biases for domains posted on Bluesky, as categorized by Media Bias/Fact Check (MBFC). The main bar chart shows the overall distribution across six categories: left, center left, center, center right, right, and extreme right. Inset charts provide a detailed breakdown of left-leaning (left of center) and right-leaning (right of center) domains.}
    \label{fig:overalldistrpol}
\end{figure}

Most of the domains spread via Bluesky are non-political. Within the domains with an associated political bias, left-leaning, specifically left-center, dominates. All but two of the top ten most spread political domains have an associated rating of left-center. Figure \ref{fig:overalldistrpol} shows the overall distribution of political domains spread via Bluesky. The bar chart on the left shows the distribution of all political domains on the website.

Over 63.4\% of all domains are classified as left-leaning. Around 18.8\% of the domains are classified as center, and 7.9\% of the domains are right-leaning. This shows that Bluesky is mostly politically homogeneous, with a majority of the domains shared having an ideology left of center.  No domains are classified by MBFC as extreme left and only 0.16\% of domains are classified as extreme right. The bar charts on the right of the plot disaggregate the overall distribution into the distribution of right- and left-leaning outlets. For both left- and right-leaning outlets, the less extreme (i.e., more central) outlets dominate (left: 63.4\%, extreme-left: 0\%; right: 7.74\%, extreme-right: 0.16\%). 

To examine the political stance of users of Bluesky, as opposed to the political stance of domains shared, we average the bias scores of all domains shared by each user. Each news outlet is assigned a political leaning score ranging from left to extreme right (assigned scores from -32 to +32). Looking at the average score of domains shared per user, 75.3\% of users are left of center (69.26\% center-left, 6.05\% left) and 4.81\% of users are right of center (4.62\% center right, 0.1\% right, 0.08\% extreme right). The remaining 19.79\% are classified as center.

Cinelli et al. \cite{cinelli2021echo} study polarization and echo chambers by examining the distribution of user opinions and comparing them to the opinions of their neighborhood. The neighborhood is defined as the set of nodes directly connected to a given node in the network. The average opinion of the neighborhood is defined as \(x_i^N \equiv \frac{1}{k_i^\rightarrow} \sum_j A_{ij} x_j\) where $k_i^\rightarrow$ is the out-degree of node i, \(A_{ij}\) is the adjacency matrix of the analysed network, and \(x_j\) is the opinion of neighbor \(j\). The study uses the followership network as the basis of their analysis. We replicate their analysis of various social media sites, to investigate polarization on Bluesky, both with the interaction network of users (replying with a comment), and followership network.

\begin{figure}[!ht]
    \centering
    \includegraphics[width=\columnwidth]{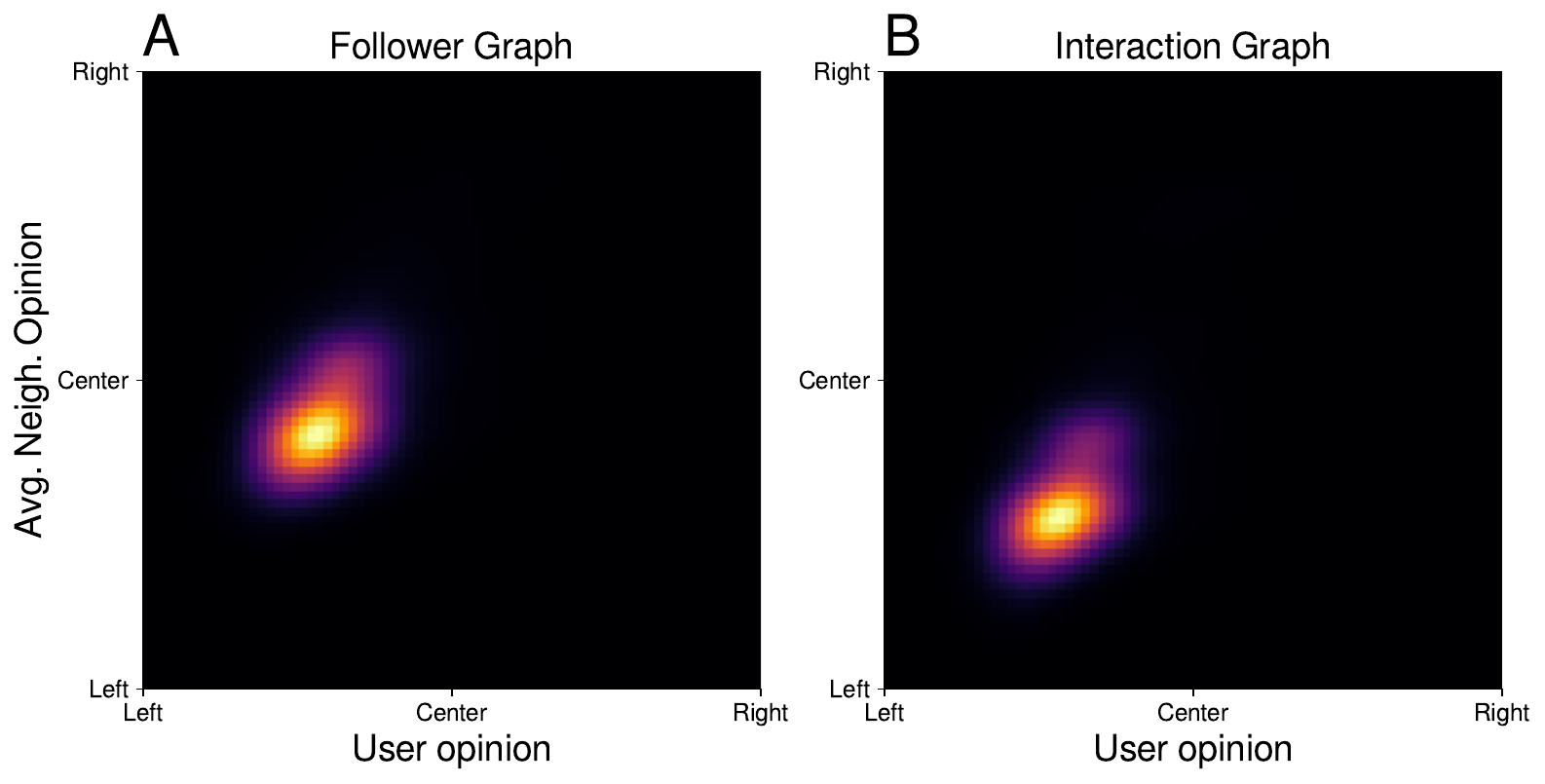}
    \caption{Heatmap of Political leaning of Users and the average of their neighborhood across the directed followership network (\textbf{A}) and the directed network of replies (\textbf{B}). Lighter areas indicate a higher density of users. Political leaning is calculated as the average political leaning of the URLs shared by a user. We exclude all users that have less than five neighbors or five posts. 
    In total, 43.074 users have shared political domains at least five times and have at least five neighbors who have done so too and are thus included in the graphic }
    \label{fig:political_opinion_neighbourhood}
\end{figure}

Figure \ref{fig:political_opinion_neighbourhood} displays the averaged political domain biases per user, as classified by MBFC, and juxtaposes them against the average biases of their neighborhood of users. Users are only included in the figure if they have shared at least five domains with a political bias and additionally have at least five neighbors who have done so too. This restricts our analysis to 43 thousand users. The distribution of user biases in the interaction and follower graph are generally very similar with a unimodal peak between center and left. This confirms our previous findings that the vast majority of users on Bluesky have a center-left political leaning. Interestingly, the distribution of the average neighborhood bias of a users is more left leaning, less central, and more similar to the leaning of the seed user in the interaction graph than in the followership graph. This indicates that while users may follow a relatively more diverse range of political perspectives, their actual interactions are more politically homogeneous, highlighting a discrepancy between passive following behavior and active engagement patterns on the platform.

To investigate the polarization of opinions on a specific issue on Bluesky, we establish a corpus of posts related to the Israel-Palestine conflict and train a machine learning model to predict the stance of a post towards the conflict. Training details, test-set classification reports, and details on data labeling and querying are available in the Materials and Methods section of the manuscript. 

\begin{figure}[!ht]
    \centering
    \includegraphics[width=\columnwidth]{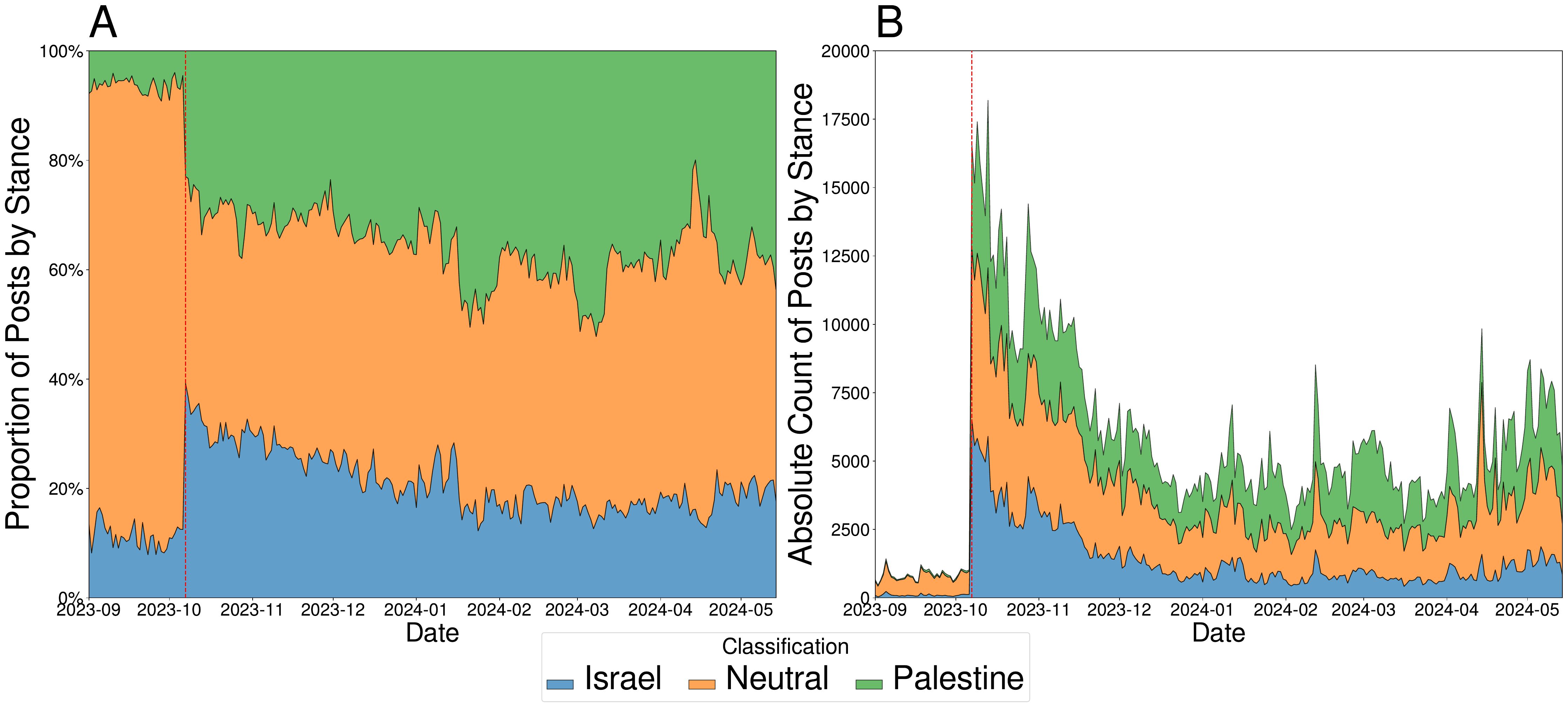}
    \caption{Distribution of Posts by Stance on the Israel-Palestine Conflict Over Time. (\textbf{A}) Daily proportions of posts, with the y-axis representing the percentage of total posts for each stance: neutral (orange), pro-Palestine (green), and pro-Israel (blue). Notably, neutral posts predominate until October 7, 2023, when a marked shift occurs towards more polarized views following the onset of the latest conflict. Over the subsequent months, pro-Palestine posts gradually outnumber pro-Israeli voices by January 2024. In the month preceding the attack on October 7th, the percentage of neutral posts dropped from 82.86\% to 37.98\%. While in October, pro-Israel voices outnumbered pro-Palestinian voices (33.01\% vs 28.99\%), in the final month of the observation period, only 20.74\% of messages were pro-Israel, compared to 39.00\% of messages containing a pro-Palestinian sentiment.
    (\textbf{B}) Absolute number of posts per day, with a significant spike in discussion beginning on October 7, 2023, followed by a stabilization in early 2024. This graph captures the fluctuations and trends in discourse surrounding the conflict from July 2023 to May 2024.}
    \label{fig:PalestineOverTime}
\end{figure}

Figure \ref{fig:PalestineOverTime}\textbf{A} shows the proportion of posts by stance per day. The y-axis represents the percentage of total posts for each stance, spanning from 0\% to 100\%. The x-axis covers the date range from July 2023 to early May 2024. The graph color-codes the posts: orange indicates neutral posts, green represents pro-Palestine posts, and blue signifies pro-Israel posts. The proportions of each stance change over time, with a notable dominance of neutral stances before October 7, 2023. On and after this date, there is a visible shift in the distribution of stances. Following the attacks on Israel, the percentage of neutral posts shrinks with an increase in both the number of Pro-Palestinian and Pro-Israel stances. Over the course of the following ten months, the percentage of Pro-Palestinian messages increases steadily, reaching the absolute majority of posts in January 2024.

Figure \ref{fig:PalestineOverTime}\textbf{B} displays the absolute count of posts by stance per day. Similar to Figure \ref{fig:PalestineOverTime}\textbf{A}, it uses the same color coding for each stance and spans the same time period on the x-axis. The y-axis, however, measures the count of posts, ranging from 0 to 18,000 posts per day. Prior to October 7, only a very small number of posts discussed Palestine \& Gaza. On October 7, we see a spike with a gradual decay in posts until January 2024. Since then, the number of posts per day has remained relatively stable at around 4,000 messages.

\begin{figure}[!ht]
    \centering
    \includegraphics[width=\columnwidth]{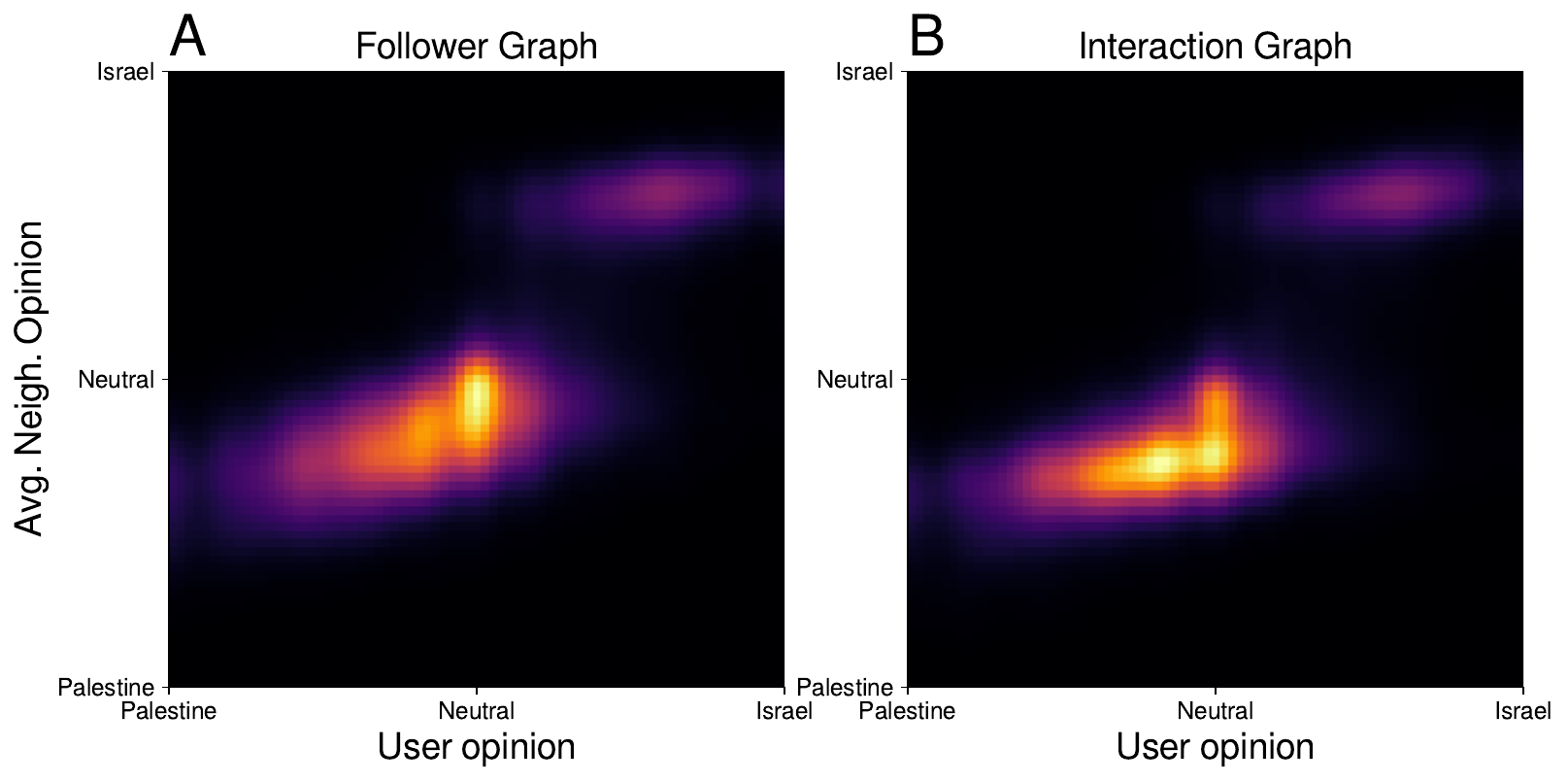}
    \caption{Heatmap of users' stance versus their neighborhood's average stance. The main plot shows the distribution across the directed followership network (\textbf{A}) and the directed network of replies (\textbf{B}). Lighter areas indicate a higher density of users. Political leaning is calculated as the average political leaning of URLs shared by a user. The figure includes 30.048 users who have shared at least five posts related to the conflict and have at least five neighbors who have done the same.}
    \label{fig:PalestineIsrael}
\end{figure}

To examine how users' stances on the Israel-Palestine conflict are distributed across the Bluesky network and how these stances relate to users' social connections, we conduct a network analysis similar to our earlier examination of general political ideology. This analysis allows us to visualize potential echo chambers or polarization specific to this issue, which may differ from the overall political leaning of the platform. We again extract all users with at least five posts indicating an opinion on the subject and average their political stances to map each user onto a one-dimensional stance. We then calculate the average stance of every user with at least five posts and five neighbors. The results are shown in figure \ref{fig:PalestineIsrael}. Both networks showcase a similar neighborhood opinion graph with two distinct clusters. The majority of users are concentrated in two areas: a larger, more diffuse cluster spanning from Neutral to Pro-Palestine stances and a smaller, more compact cluster in the Pro-Israel region. A clear diagonal trend from bottom-left to top-right indicates users tend to connect with others holding similar views. 

Although Bluesky predominantly displays a left-leaning political bias, this does not reflect uniformity of opinions on all subjects. The analysis of discussions on the Israel-Palestine conflict reveals a spectrum of stances within the platform. This range of perspectives highlights that political homogeneity does not necessarily dictate consensus on specific issues with polarized debates.

\section*{Discussion}
Bluesky, for all its innovative features, is a social media site that resembles larger and older sites in almost all of its network features. Our analysis reveals patterns of clustering and small-world properties analogous to those observed in platforms like Twitter (now X). Our investigation into Bluesky's user composition reveals both homogeneity and diversity. While the platform exhibits a predominantly left-leaning user base in terms of broad political orientation, similar to some other small platforms, it demonstrates significant diversity in opinions on specific issues such as the Israel-Palestine conflict. Even within seemingly homogeneous platforms, there is potential for diverse discourse on particular topics. 
Future work could investigate whether such polarization of opinions on specific issues is indicative of a healthy dialogue in the marketplace of ideas or if it is driven by affective polarization and political sectarianism, which emerged in platforms such as Twitter and Facebook\cite{finkel_political_2020}.
Our findings contribute to our understanding of how user bases form and evolve on emerging social media platforms and highlight the importance of investigating polarization by looking beyond a simple left-right spectrum. The creation of feeds has been taken up by users with enthusiasm with almost forty thousand choices present for users. However, only a small minority of users have liked a feed. Bluesky enables researchers to answer old questions with a novel treasure trove of data that could contribute to a range of scientific open questions.

\section*{Materials and Methods}

\subsection*{Data}
The data for this study consists of the complete repositories of five million users on Bluesky, each containing all associated user profiles and actions. Due to the decentralized nature of the platform, Bluesky repository data is accessible to any person with the ID of a user. For each user, we first queried the centralized directory of user DIDs (DID PLC directory) \cite{did_plc_directory} to get the personal data server address (PDS) of the user's repository. Given the DID of the user and the address of the PDS, the repository data can then be queried. The code to download user repositories can be found at \url{https://github.com/dorianquelle/BlueskyTopology/blob/main/Code/PythonScripts/download_repos.py}.

An initial seed of 5.28 million IDs of Bluesky users was posted by a Bluesky contributor on the 26th of March 2024 \cite{jaz_bsky_social_2024}. Given this initial seed, we extracted all data for all users who remained active in the dataset and subsequently checked for any users referenced in the downloaded data. We repeated this procedure until no new users were found. The final dataset was cleaned and stored in a database containing individual tables for likes, follows, posts, reposts, blocks, and feed creations.

Table \ref{tab:data_summary} contains a summary of the number of rows for each of the SQL tables. In addition, the column ``unique users'' contains the number of users who have authored an action (like, post, repost, follow, feed) in the table. 

\begin{table}[!ht]

\centering
\caption{Number of rows and unique users per table in the dataset.}
\label{tab:data_summary}
\begin{tabular}{lrrrrrr}
\toprule
 & \textbf{Blocks} & \textbf{Follows} & \textbf{Posts} & \textbf{Likes} & \textbf{Reposts} & \textbf{Feeds} \\
\midrule
Total Rows & 8,357,905 & 149,650,293 & 206,346,303 & 771,009,280 & 81,655,778 & 39,639\\
Unique Users & 479,427 & 3,844,491 & 2,339,109 & 2,581,744 & 1,197,330 & 18,352\\
\bottomrule
\end{tabular}

\end{table}

\subsection*{Interactions Networks}
\paragraph{Followership Network} An edge connects a user to another if they follow that user. In contrast to Facebook, connections between users are not reciprocal, meaning that the network is directed. Followership relations are not transient but persistent. An edge between two users remains until it is removed. Following another user generally indicates an interest in the content that they post, as their content will be shown to them in the user's main feed. Alternatively, followership could be an indicator for a social relationship outside of Bluesky, meaning that the two users are more likely to share socio-demographic features.

\paragraph{Replies Network} An edge connects two users if one responds to another in the same thread with a comment. This network is also directed, but in contrast to the followership network, it is not persistent. Responding can indicate an overlap of the thematic interests of two users---but does not imply agreement.

\paragraph{Repost Network}: Two users are connected by an edge if a user reposts a post by another user. A repost on Bluesky is equivalent to a retweet on Twitter. The network is non-persistent and directed. A repost indicates an interest in the post of another user \cite{boyd2010tweet}. Additionally, the user is willing to share the content with their own followers \cite{kleppmann2024bluesky}.

\paragraph{Likes Network} An edge in the Likes Network indicates whether a user liked a post from another user. The network is directed and non-persistent. Liking a post of another user indicates interest in the topics posted by the user \cite{levordashka2016s}. In contrast to the Repost Network, the post will not be shown to the user's followers.

\subsection*{Stance Detection for the Israel-Palestine Conflict}
\paragraph{Israel Palestine Term Extraction \& Data Labelling}
First, we extracted all posts containing the keywords ``Israel'', ``Palestine'' and their translations into languages present in the dataset (Arabic, English, French, German, Greek, Hebrew, Italian, Japanese, Korean, Persian, Russian, Ukrainian, Azerbaijani, Danish, Dutch, French, Finnish, Hungarian, Indonesian, Kazakh, Chinese, Norwegian, Portuguese, Romanian, Slovene, Spanish, Swedish, Tajik, Turkish). Subsequently, we calculated the mutual information of all uni-, bi-, and tri-grams from the dataset with any of initial seed terms, filtering for a minimal number of occurrences of above 50 times. We then manually reviewed the top 100 n-grams, ranked by their mutual information, extracted for each of the initial seed terms and selected all n-grams directly related to the conflict. Lastly, to prevent any biases across languages, we translated all selected n-grams into all the initial languages to ensure that no variation in n-grams included bias the distribution of stance scores. A full list of all terms with at least 1.000 exclusive posts (posts not added by any other n-grams), can be found in table \ref{tab:QueryTerm} in the appendix. We then queried the database for any of the retrieved n-grams and created a dataset of 1.3 Million posts related to the conflict. From this subset, we manually annotated a random sample of 1,000 posts. Each post was labeled as Pro-Israel (1), Neutral (0), or Pro-Palestine (-1) based on the stance expressed in the content. An additional set of 1.000 posts was labeled by crowdworkers via \url{Appen.com}. Crowdworkers were presented with posts from Bluesky and prompted with the following question. ``Determine the stance (favor Israel, favor Palestine, or neither) of the author towards the Israel/Palestine conflict from each given social media post.'' Each participant was given examples for each of the categories and additionally tested on ten quality assurance questions. The average agreement between annotators was 71.23\%. 
We additionally removed any posts where no majority opinion among annotators emerged.  

\paragraph{Stance Prediction}
Stance prediction involves automatically determining the position or attitude expressed in a piece of text toward a specific target or topic. We use the multilingual transformer-based language model XLM-RoBERTa large \cite{xlmroberta}, which is well-suited for the stance prediction task due to its ability to capture cross-lingual semantic information. The data is preprocessed by tokenizing the text using the XLM-RoBERTa tokenizer, with a maximum sequence length of 128 tokens. The model is trained on an A100 GPU and evaluated on a hold-out test set. The final training configuration is reached via fine-tuning had a dropout-probability of 0.14, a learning rate of 2.2$\times$ 10\textsuperscript{-05} and a weight decay of 0.001. In addition, we freeze the first 21 (out of 24) layers of the model to reduce the chance of overfitting and accelerate training. The best model is finally determined by maximizing the Macro F1 score. Predictions are obtained by applying the trained model to the tokenized test data and selecting the class with the highest probability. Classification metrics, including precision, recall, and F1-score, are reported for each class in table \ref{tab:model_comparison}. The model significantly outperforms random guessing and a simple benchmarking naive Bayes model, which achieves a macro F1 score of 0.47, shown in table \ref{tab:naive_bayes_results}.

\begin{table}[!ht]
\centering
\caption{Test Set Classification Performance}
\label{tab:model_comparison}
\begin{tabular}{lcccc}
\toprule
\textbf{Class} & \textbf{Precision} & \textbf{Recall} & \textbf{F1-Score} & \textbf{Support} \\
\midrule
\textbf{Palestine} & 0.623 & 0.716 & 0.667 & 60 \\
\textbf{Neutral} & 0.712 & 0.602 & 0.653 & 78 \\
\textbf{Israel} & 0.595 & 0.647 & 0.620 & 34 \\
\midrule
\textbf{Accuracy} & & & & 0.653 \\
\textbf{F1 Score} & & & & 0.646 \\
\bottomrule
\end{tabular}
\end{table}

\begin{table}[!ht]
\centering
\caption{Test Set Classification Report for Naive Bayes}
\label{tab:naive_bayes_results}
\begin{tabular}{lcccc}
\toprule
\textbf{Class} & \textbf{Precision} & \textbf{Recall} & \textbf{F1-score} & \textbf{Support} \\
\midrule
\textbf{Palestine} & 0.471 & 0.817 & 0.598 & 60 \\
\textbf{Neutral} & 0.565 & 0.333 & 0.419 & 78 \\
\textbf{Israel} & 0.500 & 0.324 & 0.393 & 34 \\
\midrule
\textbf{Accuracy} & & & & 0.500 \\
\textbf{F1 Score} & & & & 0.470\\
\bottomrule
\end{tabular}
\end{table}

\subsection*{Network Metrics}
\paragraph{Power Law Exponent}
To estimate the exponent of the power-law degree distribution, we employed a maximum likelihood estimation method \cite{newman2005power}. We considered only degrees $k \geq k_{min}$ where $k_{min} = 5$, and calculated the exponent $\alpha$ using the formula:
$$\alpha = 1 + n \left[ \sum_{i=1}^n \ln \frac{k_i}{k_{min}} \right]^{-1}$$ where $n$ is the number of nodes with degree $k \geq k_{min}$, and $k_i$ are the observed degrees that meet this criterion.

\paragraph{Clustering Coefficient}
The clustering coefficient provides insights into the local structure of the network, indicating how likely it is for nodes to form tightly connected groups. A high clustering coefficient suggests a network with many triangles. In the context of social networks, the clustering coefficient can be intuitively understood as the probability that two of your friends are also friends with each other. More formally, it measures the likelihood of triadic closure in the network. The clustering coefficient for a directed graph is defined by considering the possible directed triangles\cite{fagiolo2007clustering}. 

For each node $i$, the local clustering coefficient $C_i$ is computed as:
$$C_i = \frac{\text{number of directed triangles passing through } i}{\text{number of possible directed triangles passing through } i}$$
Which is calculated as:

$$C_{i} = \frac{(A + A^T)_{ii}^3}{2(d_{i}^{\text{tot}}(d_{i}^{\text{tot}} -1) - 2d_i^{\leftrightarrow})}$$

where $(A + A^T)_{ii}^3$ is the number of directed triangles including node $i$, $d^{\text{tot}}$ is  the sum of in and out degrees of node $i$. Lastly, $d_i^{\leftrightarrow}$ is the reciprocal degree of node $i$, i.e. the number of nodes j for which both an edge $i\rightarrow j$ and an edge $j\rightarrow i$ exist. For the entire network, the clustering coefficient $C$ is computed as the average of all individual local clustering coefficients. We normalize the clustering coefficient by creating a randomized configuration-model graph with the same degree in- and out-degree sequences than the empirical graphs. In the random graph, the in- and out-degree of each node is pre-defined, but nodes are randomly connected. The normalized clustering coefficient is defined as  $CC_{\text{norm}} = \frac{CC}{CC_{\text{rand}}}$\cite{watts1998collective}. 
\paragraph{Network Density}
Network density measures the proportion of potential connections in a network that are actual connections. It provides insight into the overall connectivity and compactness of the graph. For a directed graph, the network density $D$ is defined as:
$$D = \frac{m}{n(n-1)}$$
where $m$ is the total number of directed edges in the graph, and $n$ is the total number of nodes. 

\paragraph{Average Shortest Path Length}
The average shortest path length is a measure of the efficiency of information or traffic flow within a network. It quantifies the average number of steps along the shortest paths for all possible pairs of network nodes. It is a significant indicator of the 'small-world' characteristic of a network. A network exhibits the 'small-world' characteristic if ``any two individuals in the network are likely to be connected through a short sequence of intermediate acquaintances'' \cite{kleinberg2000small} \cite{watts1998collective}. This metric indicates the ease with which information spreads across the network and is a key factor in the analysis of network efficiency and connectivity. We provide an approximate value for the average shortest path in each network by randomly sampling 50 thousand pairs of nodes in the graph and calculating the mean distance between the nodes.

\paragraph{Degree Distribution}
The degree distribution of a network describes the relative frequency of nodes with different degrees within the graph. In social media networks, this distribution is often heavy-tailed, with a power-law-like shape, indicating that while most users have few connections, a small number of users (hubs) have a disproportionately large number of connections, but also the absence of clear scale separations between users based on connections \cite{raban2009statistical} \cite{centola2015social}.

\section*{Data Availailibility Statement}
In accordance with the decision by the UZH PhF Ethics Committee, the raw data can not be shared as it contains potentially identifying information. However, we provide a code (see Data section in the Materials and Methods) that can be used to download all the data necessary to reproduce our results directly from the Bluesky API. 
\section*{S1 Appendix}
\begin{table}[ht]
\centering
\begin{tabular}{llll}
\toprule
Term & Translation & Total Posts & Exclusive Posts \\
\midrule
israel & -- & 632,737 & 274,679 \\
gaza & -- & 336,061 & 126,116 \\
hamas & -- & 233,000 & 82,169 \\
palestine & -- & 127,002 & 62,032 \\
jewish & -- & 92,716 & 55,697 \\
antisemitism & -- & 59,499 & 35,680 \\
jews & -- & 71,147 & 34,178 \\
rafa & -- & 45,856 & 29,255 \\
Israel\textsuperscript{*} & Israel & 51,748 & 28,096 \\   
zionis & Zionism & 58,451 & 27,168 \\
palestinians & -- & 90,392 & 26,539 \\
netanyahu & -- & 49,334 & 17,871 \\
judar & Jews & 14,115 & 13,575 \\
Israel\textsuperscript{†} & Israel & 15,577 & 13,277 \\  
juden & Jews & 16,453 & 9,252 \\
Israel\textsuperscript{‡} & Israel & 11,696 & 9,097 \\    
israël & Israel & 22,843 & 8,175 \\
Palestine* & Palestine & 23,045 & 7,777 \\  
jüdisch & Jewish & 12,058 & 7,324 \\
sionism & Zionism & 6,364 & 5,583 \\
judaism & -- & 9,554 & 4,616 \\
palestina & Palestine & 8,878 & 4,435 \\
ebrei & Jews & 4,256 & 4,047 \\
Hamas\textsuperscript{*} & Hamas & 12,883 & 3,190 \\      
Palestine\textsuperscript{‡} & Palestine & 5,845 & 3,149 \\   
Hamas\textsuperscript{†} & Hamas & 4,781 & 3,034 \\       
west bank & -- & 17,698 & 2,870 \\
october 7 & -- & 11,418 & 2,808 \\
juif & Jewish & 3,834 & 2,624 \\
antisémitisme & Antisemitism & 3,360 & 2,545 \\
Israel\textsuperscript{§} & Israel & 3,172 & 2,347 \\     
palästina & Palestine & 6,769 & 2,300 \\
Israel\textsuperscript{¶} & Israel & 3,007 & 2,251 \\     
palästinenser & Palestinians & 7,995 & 2,004 \\
Hamas\textsuperscript{¶} & Hamas & 3,022 & 1,960 \\       
israélien & Israeli & 6,276 & 1,952 \\
from the river to the sea & -- & 5,485 & 1,894 \\
Netanyahu\textsuperscript{*} & Netanyahu & 4,581 & 1,593 \\    
Gaza Strip\textsuperscript{*} & Gaza Strip & 4,742 & 1,427 \\  
sionisti & Zionists & 1,376 & 1,241 \\
gazze & Gaza & 1,643 & 1,207 \\
Israel\textsuperscript{¶} & Israel & 1,765 & 1,126 \\     
joden & Jews & 1,599 & 1,021 \\
\bottomrule
\end{tabular}
\caption{%
Selected query terms used to identify posts related to the Israel-Palestine conflict on Bluesky. 
Terms with at least 1,000 exclusive posts (posts not matched by any other n-gram), derived from an initial set of seed terms (``Israel,'' ``Palestine,'' and their translations) through mutual information calculation and manual review. Terms are listed in descending order of exclusive posts.
\\[0.5em]
\footnotesize
\begin{tabular}{@{}l@{}}
*~Transliterated from Arabic. \\
†~Transliterated from Hebrew. \\
‡~Transliterated from Japanese. \\
§~Transliterated from Korean. \\
¶~Transliterated from Russian.
\end{tabular}
}
\label{tab:QueryTerm}
\end{table}

\section*{Acknowledgements}
The authors would like to thank Ilya (Marshal) Siamionau for his technical support with the Bluesky API. We also thank Frederic Denker and Matteo Cinelli for their valuable inputs.

\end{document}